# Enhancing Solar Thermal Energy Conversion with Silicon-cored Tungsten Nanowire Selective Metamaterial Absorbers

Jui-Yung Chang,[1,2,3] Sydney Taylor,[1] Ryan McBurney,[1] Xiaoyan Ying,[1] Ganesh Allu,[2] Yu-Bin Chen,[3,4] and Liping Wang[1,*]

[1] School for Engineering of Matter, Transport & Energy
Arizona State University, Tempe, AZ 85287, USA
[2] Department of Mechanical Engineering
National Chiao Tung University, Hsinchu City 300, Taiwan
[3] Department of Power Mechanical Engineering
National Tsing Hua University, Hsinchu City 300, Taiwan
[4] Department of Mechanical Engineering
National Cheng Kung University, Tainan 70101, Taiwan

## Summary

This work experimentally studies silicon-cored tungsten nanowire selective metamaterial absorber to enhance solar-thermal energy harvesting, simply fabricated by conformally coating a thin tungsten layer onto a commercial silicon nanowire stamp. Scanning electron microscopy is used to characterize the morphology change before and after the tungsten deposition, while optical spectroscopy is carried out to measure the spectral absorptance (or emittance) in the broad wavelength range from solar spectrum to infrared. It is shown that the tungsten nanowire absorber exhibits almost the same total solar absorptance around 0.85 as the silicon nanowire but with greatly reduced total emittance down to 0.18, which could significantly suppress the infrared emission heat loss. A lab-scale solar-thermal test apparatus is used to measure the solar-thermal efficiency of the tungsten nanowire absorbers from 1 up to 20 suns, experimentally demonstrating the improved performance due to excellent spectral selectivity over the silicon nanowire and black absorbers. With detailed heat transfer analysis, it is found that the tungsten nanowire absorber achieves an experimental efficiency of 41% at temperature 203°C during the solar-thermal test with stagnation temperature 273°C under 6.3 suns. It is projected to reach 74% efficiency at same temperature 203°C with stagnation temperature of 430°C for practical application without parasitic radiative losses from side and bottom surfaces, greatly outperforming the silicon nanowire and black absorbers. The results would facilitate the development of novel metamaterial selective absorbers at low cost for highly-efficient solar thermal energy systems.

[*] Corresponding author. Email: liping.wang@asu.edu, Phone: 1-480-727-8615

# Introduction

Within last decade, nanostructured metamaterials have become an attractive topic in the fields of radiative heat transfer for thermal energy harvesting (Bello and Shanmugan, 2020; Jin et al., 2016; Khodasevych et al., 2015; Woolf et al., 2018; Xu et al., 2020) and radiative cooling (Raman et al., 2014; Zhai et al., 2017). For harvesting solar energy to heat, spectrally selective absorbers with high solar absorption and low infrared emission are highly desired for efficient energy conversion, and many metamaterial selective solar absorbers have been designed and experimentally demonstrated recently based on multilayer (Khoza et al., 2019; Thomas et al., 2017; Wang et al., 2018), periodic tungsten convex or concave gratings (Jae Lee et al., 2014; Wang et al., 2015; Wang and Wang, 2013), nickel nanopyramids and tungsten nanowires/doughnuts (Behera and Joseph, 2017; Li et al., 2015; Tian et al., 2018), nanoporous or nanoparticle composite structures (Lu et al., 2016; Lu et al., 2017; Prasad et al., 2018). Due to their submicroscale feature sizes, advanced fabrication techniques such as electron-beam lithography and focused-ion beam were usually needed for fabricating these metamaterial structures (Wang et al., 2015), which are expensive with low throughput prohibiting their large-area applications. Some cost-effective methods such as direct printing (Han et al., 2016) and interference lithography (Li et al., 2015) in combination with additional thin film deposition and selective etching processes were also successfully demonstrated for fabricating cm-scale selective metamaterial absorbers. Simpler and more cost-effective methods are still in need for large-area manufacturing of selective metamaterial solar absorbers. These experimentally demonstrated metamaterial solar absorbers exhibit the typical periodicity and feature size around a few hundred of nanometers, which can be feasibly fabricated with deep UV projection or stepper photolithography (Enrico et al., 2019; Greiner et al., 2006; Park and Lee, 2016). In fact, diffraction gratings for optical devices and stamps for nanoimprinting made of nanopatterned silicon wafers fabricated by deep UV projection lithography along with selective chemical or plasma etching (Opachich et al., 2015; Sanjay et al., 2014) have been commercialized in large area with controllable shape and geometries (Lightsmyth).

Moreover, most of reported studies only focused on the fabrication and characterization of the optical and radiative thermal properties to demonstrate the spectral selectivity of metamaterial solar absorbers with theoretical prediction of their solar-thermal performances. To bridge the gap between material-level researches to device applications, lab-scale solar-thermal characterization must be implemented to evaluate the energy conversion performance of developed metamaterial absorbers under different solar concentrations and experimentally ensure the expected performance before large-are manufacturing and more comprehensive outdoor field tests. From an outdoor field test under unconcentrated solar irradiation, a peak temperature of 225°C for their semiconductor-based multilayer selective solar absorber was experimentally obtained inside a vacuum chamber with a glass lid (Thomas et al., 2017). Lab-scale photo-thermal conversion experiment with one-sun illumination was conducted for Cr grating structure that reached a high temperature of 80.7ºC (Seo et al., 2019). A novel lab-scale solar-thermal test platform with a broadband solar simulator was successfully developed to characterize the performance of a fabricated metafilm selective solar absorber under multiple solar concentrations up to 20 suns along with detailed heat transfer analysis (Alshehri et al.). So far systematic lab-scale solar-thermal characterization of nanostructured metamaterial selective solar absorbers under variable solar irradiation is still barely demonstrated.

In this study, we present the simple fabrication, optical characterization, and lab-scale solar-thermal tests of selective metamaterial solar absorber made of silicon-cored tungsten nanowires to experimentally demonstrate the improved performance in converting solar energy to heat under multiple solar concentrations due to its excellent spectral selectivity. The fabrication processes will be discussed first along with morphology characterizations, followed by the spectroscopic measurements of the optical and radiative properties in the broad spectrum from solar to infrared. Numerical simulation is then conducted to elucidate the physical mechanisms of high solar absorption and low infrared emission. Lastly, lab-scale solar-thermal tests under variable suns are carried out to evaluate and demonstrate its superior conversion performance in comparison to the bare silicon nanowire and black absorbers, which is validated by detailed heat transfer analysis.

## Results and Discussion

**Sample fabrication and characterization**

The silicon-cored tungsten nanowire (WNW) selective metamaterial absorber was fabricated by conformally sputtering a thin tungsten layer (Lesker PVD75) at a depositing rate of 1 Å/s onto a commercial 2D silicon nanowire (SiNW) stamp (LightSmyth Technologies, S2D-18B2-0808-350-P, 1-10 Ω·cm resistivity, single-side polished, 8×8 mm$^2$, 675 μm thickness), as illustrated in Fig. 1(a). Tungsten was selected to be the coating material because of its excellent high-temperature stability, high absorption in the visible and near-infrared, and very low emission in the mid-infrared regime. The fabricated WNW sample is visibly black as seen from the photo in Fig. 1(b). Scanning electron microscope (SEM) was used to characterize the geometry of the SiNW sample before and after the tungsten coating, with the top-view images respectively shown in Figs. 1(c) and 1(d). Clearly the period of the nanowires is $P$ = 600 nm, which does not change with the tungsten deposition, while the nanowire diameter increases from $D_{in}$ = 275 nm to $D_{out}$ = 350 nm, indicating 37.5-nm-thick tungsten was conformally coated onto the silicon nanowires. With the groove depth of 350 nm (provided by the vendor) for the nanopatterned silicon stamp, the silicon-cored WNW was estimated to have the height of $H$ = 350 nm due to conformal sputtering. Note that another same sample of the bare SiNW nanostamp without the tungsten coating was used along with the fabricated Si-cored WNW metamaterial sample as solar absorbers for the comparison studies of the spectral radiative properties and solar thermal tests, which are to be discussed in following sections. In order to ensure the opaqueness, 200-nm-thick aluminum was later sputtered onto the backsides of the samples (denoted as WNW-Al and SiNW-Al) at 1 Å/s.

**Spectral and total radiative properties**

Figure 2(a) presents the measured spectral (near-normal) reflectance ($R_\lambda$) of the SiNW and silicon-cored WNW samples in the broad wavelength range 0.4 μm to 25 μm. Experimental details about optical measurements can be found in Supplemental Information. Within the visible to and near-infrared, both SiNW and WNW samples exhibit comparable low reflectance with a peak value around 0.30. However, the reflectance of the WNW sample increases abruptly at wavelengths beyond 2 μm, reaching $R_\lambda$ = 0.75 around $\lambda$ = 5 μm, at which the bare SiNW only has a reflectance of 0.30. Note that for a selective solar absorber, the infrared reflectance should be as high as possible to minimize thermal emission losses. The much higher infrared reflectance of the WNW sample indicates a potential better solar-thermal conversion than the bare SiNW, thanks to the highly reflective tungsten coating in the infrared. In practical applications, the solar absorber is expected to be opaque such that the thermal emission from the hot heat transfer fluids underneath will not penetrate to lose the heat. In fact, the spectral (normal) transmittance measurement reveals that the bare SiNW sample still has about up to 15% transmission from 1 µm to 3.5 µm in the near-infrared because of the low silicon doping, while the WNW sample along with the 37.5-nm-thick bare tungsten film deposited on lightly doped silicon wafer show only about 2% transmission in the same wavelengths (Fig. S1a in the Supplemental Information). Due to the bandgap absorption and large thickness of silicon wafer, the transmission at wavelengths $\lambda < 1$ µm for both WNW and SiNW samples is basically zero. Therefore, 200-nm-thickness Al film was sputtered at a rate of 1 Å/s onto the backsides of both samples to ensure the complete opaqueness, which was confirmed by the zero transmittance from the measurement. The broadband reflectance measurement further verifies the effectiveness of the Al backside coating, in particular for bare SiNW, whose infrared reflectance was increased from 0.35 to 0.55 at the wavelength $\lambda = 25$ µm. On the other hand, the measured spectral reflectance of the WNW barely changes after Al backside coating within the entire spectrum of interest, and remains as high as 0.85 in the infrared.

With the Al backside coating, it is reasonable to calculate the spectral absorptance simply by $\alpha_\lambda = 1 - R_\lambda$ and the spectral emittance $\varepsilon_\lambda = \alpha_\lambda$ according to Kirchhoff's law. Note that in principle the spectral reflectance should be the hemispherical one for calculating the absorptance from energy balance. However, for periodic structures under normal incidence, only specular reflection occurs at wavelengths longer than its period with higher diffraction orders to be evanescent ones according to the Bloch-Floquet equation for gratings (Wang et al., 2015; Zhang, 2007). Therefore, the measured spectral specular reflectance can be directly used for calculating the spectral normal absorptance at most spectrum with wavelength $\lambda > 0.6$ µm for both WNW-Al and SiNW-Al absorbers, while it is approximated as the hemispherical one for simplicity here at the shorter spectrum 0.4 µm $\lambda < 0.6$ µm.

Figure 2(b) plots the spectral (normal) absorptance $\alpha_\lambda$ (or emittance $\varepsilon_\lambda$) for both SiNW-Al and Si-cored WNW-Al absorbers calculated from the spectral reflectance measurements in the wavelengths from 0.4 µm to 25 µm, where the spectral solar irradiance $G_{AM1.5}(\lambda)$ from air mass 1.5 Global tilt (shaded orange) (Air Mass 1.5 Spectra) and spectral blackbody emissive power $E_b(\lambda,T)$ at $T = 400°C$ (shaded green) (Modest, 2013) both normalized to the peak spectral emissive power of 0.178 W/cm²/µm are also shown. Clearly the WNW-Al sample has absorption higher than 70% within the entire solar spectrum thanks to several absorption peaks observed at $\lambda$ = 2.44 µm, 1.89 µm, 1.28 µm, and 0.60 µm, while its emittance is as low as 0.15 in the infrared, suggesting strong solar absorption and excellent suppression of infrared thermal emission. Numerically modeled spectral absorptance of the WNW-Al absorber verified the measurement (Fig. S1b in the Supplemental Materials), showing similar low emittance in the infrared down to 0.05 at $\lambda$ = 25 µm possibly due to theoretical material properties of pure tungsten and similar high absorption within solar spectrum with multiple peaks, whose mechanisms are to be explained next. On the other hand, the SiNW-Al sample has comparably high absorptance in the solar spectrum with major peaks at $\lambda$ = 1.1 µm and 0.60 µm, but its infrared spectral emittance is greater than 0.6 at wavelengths $\lambda$ < 10 µm and is still as high as 0.45 at $\lambda$ = 25 µm, indicating higher thermal emission loss than the WNW sample. Note that the minor absorption peak around $\lambda$ = 15 µm is due to intrinsic optical phonon absorption of silicon. Undoubtedly the 37.5-nm tungsten front coating significantly reduced the spectral emittance of the bare SiNW by up to 0.52 in the wavelength range 3 µm < $\lambda$ < 10 µm where the most of blackbody emission spectrum at 400°C is located, and excellent spectral selectivity is experimentally demonstrated with the silicon-cored WNW absorber.

Note that a black surface at 400°C dissipates heat in thermal emission (i.e., 11.6 kW/m$^2$ from $E_b = \sigma T^4$) about 11.6 times of the total solar irradiance (i.e., $G_{AM1.5} = 1$ kW/m$^2$) without concentration. Therefore, it is crucial to lower the infrared emittance for minimizing the heat loss from thermal emission for highly efficient solar thermal energy harvesting. For better quantitative comparison, the total solar absorptance $\alpha_{solar}$ is respectively 0.852 and 0.855 for the WNW-Al and SiNW-Al absorbers calculated from measured spectral absorptance $\alpha_\lambda$ and air mass 1.5 global tilt solar irradiation data based on the following equation

$$\alpha_{solar} = \frac{\int_{0.4\ \mu m}^{4\ \mu m} \alpha_\lambda G_{AM1.5}(\lambda) d\lambda}{\int_{0.4\ \mu m}^{4\ \mu m} G_{AM1.5}(\lambda) d\lambda} \tag{1}$$

On the other hand, the total emittance $\varepsilon_{th}(T)$ was calculated with the measured spectral emittance $\varepsilon_\lambda$ and spectral blackbody emissive power $E_b(\lambda, T)$ as

$$\varepsilon_{th}(T) = \frac{\int_{0.4\ \mu m}^{25\ \mu m} \varepsilon_\lambda E_b(\lambda, T) d\lambda}{\int_{0.4\ \mu m}^{25\ \mu m} E_b(\lambda, T) d\lambda} \tag{2}$$

As presented in Fig. 2(c), the total emittance from room temperature 20°C to 600°C is only 0.18 to 0.38 for the WNW-Al absorber but 0.66 to 0.74 for the SiNW-Al absorber, clearly indicating a large reduction of 0.36 to 0.48 by simply coating 37.5-nm tungsten onto the commercial SiNW nanostamp to achieve excellent spectral selectivity with significantly suppressed infrared emission while maintaining high solar absorption greater than 85%.

As thermal stability is another major concern for selective solar thermal absorbers to robustly operate at high temperatures, the spectral reflectance of both WNW-Al and SiNW-Al samples were measured after heating in vacuum with a maximum temperature of 375°C for 10 hrs during the solar thermal tests, which are to be discussed in detail later. As shown in Fig. 2(d), the measured spectral reflectance little changes in the broad spectral range before and after the heating, while the WNW-Al absorber exhibits no more than 8% decrease of reflectance (or increase of absorptance within the infrared from 1.5 μm to 10 μm, which is believed due to the possible oxidation of tungsten with residual oxygen atoms during solar thermal tests. This confirms the excellent thermal stability of both WNW-Al and SiNW-Al metamaterial absorbers.

**Underlying physics for spectral selectivity**

To elucidate the physical mechanisms responsible for the high absorption within the solar spectrum and low infrared emission from the WNW-Al absorber, spectral absorptance was numerically modeled which reasonably agrees with the measured one in Fig. S1(b). Details about the numerical simulation method can be seen in Supplemental Information. Here the simulated electromagnetic field distributions in the x-z cross-section for consecutive two unit cells of the nanowire array are plotted in Fig. 3 at four particular wavelengths where spectral absorptance peaks are observed. Note that the contour color indicates the magnetic field strength normalized to the incident field and the black arrows represent the electric field vectors. As shown in Fig. 3(a) for the wavelength $\lambda = 2.44$ μm, a strong magnetic field confinement is located in the vacuum spacing between two neighboring tungsten nanowires, whereas the electric field vectors indicate an electric current loop. This can be confirmed as the first harmonic mode of magnetic polariton (MP) as discussed in our previous theoretical study of tungsten nanowire solar absorbers which can be further verified by an inductor-capacitor circuit model prediction of 2.375 μm (Chang et al., 2017). At the wavelengths of 1.89 μm and 1.28 μm whose electromagnetic fields are respectively presented in Figs. 3(b) and 3(c), the enhancement is not caused by MP. In fact, without the electrical field forming a loop, the confinement is no longer classified as MP or artificial magnetic response. The main reason that cause the enhancement peaks is interference inside the substrate as well as the intrinsic absorption of tungsten. However, since the tungsten layer is thick enough to absorb or reflect most of the incident wave in the simulation, the effect of interference within the substrate is not as strong as MP which confines energy between NWs. That is, these two peaks are not as distinct as MP1. The last peak at 0.60 µm on the other hand, is cause by the well-known Wood's anomaly which can be predicted analytically at 0.59 µm wavelength with the utilization of the diffraction equation for periodic arrays and effective medium theory (Chen et al., 2006; Chen and Tan, 2010; Ho et al., 2016).

**Solar thermal test and performance analysis**

In order to experimentally demonstrate the performance of the WNW-Al and SiNW-Al samples as selective solar absorbers, lab-scale solar thermal test with a custom-built setup consisting of major components such as a 1-kW solar simulator, a 1-ft$^3$ vacuum chamber, and optical filters and mirrors, was conducted at variable concentrated solar irradiation from 1 up to 20 suns by different combinations of neutral density filters. A detailed description of the solar thermal test apparatus can be found elsewhere (Alshehri et al.) and thus will not be discussed here. As depicted in Fig. 4(a), a 2×2 mm$^2$ resistance temperature detector (RTD, OMEGA, F2020-100-B-100), which measures the absorber temperature up to 500°C, was glued onto the backside of the nanowire absorber samples by thermal paste to cover the entire surface. Three independent tests were conducted under a vacuum pressure less than 1×10$^{-3}$ Torr achieved by a turbo vacuum pump for the same sample under the same solar concentration, and the averaged absorber temperature at steady state with standard deviation less than 3% were reported here.

With eliminated convection and negligible conduction by air molecules under high vacuum, the energy balance at steady state for the absorber yields

$$Q_{\text{inc}} = Q_{\text{ref}} + \sum_{i=\text{t,b,s}} Q_{\text{rad},i} + Q_{\text{cond}} \tag{3}$$

Note that $Q_{inc}$ is the incident solar irradiation measured by a power sensor (Thorlabs, S310C) and the solar concentration factor can be calculated as $CF = Q_{inc}/(A_t G_{AM1.5})$. $Q_{ref} = (1 - \alpha_{solar})Q_{inc}$ is the reflected solar irradiation by the front absorber surface, and $Q_{rad,i} = A_i \varepsilon_i \sigma(T^4 - T_\infty^4)$ is the radiated heat loss from top, bottom and side surfaces (i.e., $i$ = t, b, s) to the vacuum chamber wall at $T_\infty = 20°C$, where $A_i$ and $\varepsilon_i$ are the area and total emittance of corresponding surfaces. While the top surface is the nanowire absorber whose temperature-dependent total emittance taken from Fig. 2(c), the side and bottom surfaces are respectively tungsten and thermal paste whose temperature-dependent total emittance were found from previous optical measurements (Alshehri et al.). $Q_{cond}$ is the conducted heat via RTD wires considered as useful heat gain during the solar-thermal test, and the experimental solar-thermal efficiency can be thereby defined as

$$\eta_{exp} = \frac{Q_{cond}}{Q_{inc}} = \frac{T - T_\infty}{R_{cond}(T) Q_{inc}} \tag{4}$$

where $R_{cond}(T)$ is the conduction resistance of the RTD wires that is dependent on the absorber temperature $T$. As $Q_{cond}$ cannot be directly measured, a commercial black absorber (Acktar) with solar absorptance of 0.998 and total emittance around 0.93 characterized previously (Alshehri et al.) was used to calibrate the temperature-dependent conduction heat transfer via the RTD wires. The steady-state temperature of the black absorber was measured during the solar-thermal test under different solar concentrations from 1.1, 4.0, 6.4 to 13.3 suns (see Fig. S2 in the Supplemental Information) for calculating $Q_{cond}(T)$ from the energy balance in Eq. (3), based on which the conduction resistance $R_{cond}(T)$ was fitted into a linear function of absorber temperature with $R_{cond} = -1.8076T + 1797.1$. As shown in Fig. 4(b), the experimental conduction resistance $R_{cond}$ or conduction heat transfer $Q_{cond}$ of the WNW-Al and SiNW-Al absorbers can be found from their measured steady-state temperatures during the solar-thermal tests according to the linear relation, from which the experimental solar-thermal efficiency $\eta_{\text{exp}}$ was calculated.

Figures 4(c) and 4(d) present the measured solar-thermal efficiency under multiple solar concentrations from the solar-thermal tests for fabricated SiNW-Al and WNW-Al absorbers, respectively. In particular, with parasitic radiative losses the WNW-Al absorber achieved experimental solar-thermal efficiencies of 47%, 41%, 37% and 31% under 1.7, 6.3, 9.8, and 20 suns, which is about absolute 5% higher than that of the SiNW-Al absorber under the same solar concentration. Considering almost the same solar absorptance, this solar-thermal efficiency enhancement directly observed from the solar-thermal test is undoubtedly due to the excellent spectral selectivity of the WNW-Al absorber with much smaller total emittance that suppressed the thermal emission loss. Note that the SiNW-Al absorber with relatively higher emittance still performs much better than the black absorber by absolute 11% higher experimental efficiency from the solar-thermal test under 6.4 suns. Theoretical efficiencies were also calculated by subtracting the radiative losses from all the surfaces and the reflected solar irradiance from the incident one as

$$\eta_{\text{theo}} = \frac{Q_{\text{inc}} - Q_{\text{ref}} - \sum_{i=t,b,s} Q_{\text{rad},i}}{Q_{\text{inc}}} \qquad (5)$$

Calculated theoretical efficiencies agree well with the experimentally measured ones with the difference less than 3% for both SiNW-Al and WNW-Al absorbers, which validates the solar-thermal test result. In particular, the modeling also suggests that the WNW-Al absorber could reach the stagnation temperatures (the highest temperature with zero efficiency or no heat gain) of 142°C, 273°C, 330°C, and 438°C under 1.67, 6.23, 9.87, and 20.44 suns, which is 30°C to 56°C higher than the those SiNW-Al absorber for the lab-scale solar-thermal experiment. As respectively shown in Figs. 4(e) and 4(f), heat transfer analysis pie charts plot the ratios of each heat transfer mode to the incident solar irradiation for the SiNW-Al and WNW-Al samples at the measured absorber temperatures under different solar concentration factors (CF). While both nanowire absorbers reflect about the same 15% (blue) of incident solar energy, the WNW-Al absorber loses 7% to 17% of total energy from its top surface (orange), compared to 22% to 35% from the SiNW-Al absorber from 1.67 to 20.4 suns, thanks to the greatly reduced thermal emittance simply with 37.5-nm tungsten coating. On the other hand, during the solar-thermal tests both samples suffer from the parasitic radiative losses from the side (yellow) and bottom (gray) surfaces, which account for 23% to 39% energy losses.

During practical applications, the radiative losses from side or bottom surfaces can be possibly eliminated by laminating the selective solar absorber on the surfaces of evacuated tubes, in which case the projected solar-thermal efficiency is

$$\eta_{\text{proj}} = 1 - \frac{Q_{\text{ref}} + Q_{\text{rad,top}}}{Q_{\text{inc}}} = \alpha_{\text{solar}} - \frac{\varepsilon_{\text{th}} \sigma (T^4 - T_\infty^4)}{CF \cdot G_{\text{AM1.5}}} \tag{6}$$

Figures 5(a) and 5(b) show the heat transfer analysis pie charts for the projected applications with both SiNW-Al and WNW-Al absorbers. Clearly the percentage of energy from the incident solar radiation that is converted into useful heat gain (green color), i.e., solar-thermal efficiency is increased to 78% to 68% with WNW-Al, or 63% to 50% with the SiNW-Al absorber, for solar concentrations from 1.6 to 20 suns at corresponding temperatures measured from the lab-scale solar-thermal tests. Figures 5(c) and 5(d) plot the projected solar-thermal efficiency as a function of absorber temperature at multiple solar concentrations for the SiNW-Al and WNW-Al absorbers, respectively. With the eliminated radiative losses from side and bottom surfaces, the efficiency for the projected application and the stagnation temperatures are greatly increased compared to those measured from the lab-scale solar-thermal test. In particular, at solar concentration of 6.3 suns, the WNW-Al absorber could reach a stagnation temperature of 430°C, which is 125°C higher than the SiNW-Al absorber, or 146°C higher than the black absorber, by virtue of the excellent spectral selectivity with the 37.5-nm tungsten coating onto the SiNW sample. Additional calculations also show that the WNW-Al absorber could achieve a stagnation temperature of 238°C under unconcentrated solar irradiation (i.e., $CF = 1$), which is more than 100°C higher than the SiNW-Al or black absorbers, and reach projected solar-thermal efficiency of 51% under the absorber temperature of 400°C at 10 suns, at which either the SiNW-Al or black absorber cannot gain any net amount of heat (Fig. S3 in the Supplemental Information).

## Conclusion

We experimentally demonstrated enhanced solar-thermal energy conversion with Si-cored tungsten nanowire selective metamaterial absorber, which was simply fabricated by conformally coating 37.5-nm tungsten onto a commercial silicon nanowire stamp along with 200-nm Al backside coating to ensure the opaqueness. Optical spectroscopic measurement clearly showed the greatly improved spectral selectivity for the WNW-Al absorber with almost the same total solar absorptance around 0.85 and much reduced total emittance in the infrared from 0.18 to 0.38 at room temperature up to 600°C. Numerical simulations elucidated the mechanisms of several spectral absorptance peaks within the solar spectrum as a result of magnetic resonance, wave interference/ intrinsic absorption, and Wood's anomaly. Lab-scale solar-thermal tests further confirmed the greatly improved solar-to-heat conversion efficiency with detailed heat transfer analysis for the WNW-Al absorber over the SiNW-Al and black ones at multiple solar concentrations. In particular, the WNW-Al absorber achieved an experimental efficiency of 41% at temperature 203°C during the solar-thermal tests with stagnation temperature 273°C under 6.3 suns, which are projected to reach 74% efficiency at same temperature 203°C with stagnation temperature of 430°C for practical application without parasitic radiative losses from side and bottom surfaces, dramatically outperforming the SiNW-Al and black absorbers. The findings would facilitate the development of novel metamaterial based selective absorbers at low cost for highly-efficient solar energy harvesting and conversion.

## Acknowledgements

This work was mainly supported by Air Force Office of Scientific Research (Grant No. FA9550-17-1-0080), and in part by National Science Foundation (Grant No. CBET-1454698), NASA Space Technology Research Fellowship (Grant No. NNX16AM63H), ASU Fulton Undergraduate Research Initiative, Ministry of Science and Technology (MOST) of Taiwan (Grant No. 108-2218-E-007-009, 109-2622-E-007-015-CC3, and 109-2636-E-009-014-), and New Faculty Startup Program at National Chiao Tung University. We would also like to thank the ASU NanoFab and Eyring Center for use of their nanofabrication and characterization facilities supported in part by National Science Foundation (Grant No. ECCS-1542160).

## Authors' Contributions

JYC, XY and LW carried out the optical measurements; ST and XY fabricated the samples with materials characterization; RM conducted the solar thermal measurement and LW analyzed the data and prepared the figures; JYC, GA and YBC performed the numerical simulation; LW conceived and supervised the project; LW and YBC secured the funding; JYC drafted the original manuscript and LW revised manuscript; all authors reviewed and approved the final manuscript.

## Declaration of Interests

The authors declare no competing financial interests.

## Supplemental Information

Methods on optical measurement and numerical simulation, and Figures S1 – S3.

## References

Acktar http://www.acktar.com/category/products/lights-absorbing-foils/ultra-diffusive (accessed 21.8.20).

Air Mass 1.5 Spectra http://rredc.nrel.gov/solar/spectra/am1.5/ (accessed 26.11.18).

Alshehri, H., Ni, Q., Taylor, S., McBurney, R., Wang, H., and Wang, L. Solar Thermal Energy Conversion Enhanced by Selective Metafilm Absorber Under Multiple Solar Concentrations at High Temperatures. Submitted *https://arxiv.org/abs/1910.07079*.

Behera, S., and Joseph, J. (2017). Plasmonic metamaterial based unified broadband absorber/near infrared emitter for thermophotovoltaic system based on hexagonally packed tungsten doughnuts. J. Appl. Phys. *122*, 193104.

Bello, M., and Shanmugan, S. (2020). Achievements in mid and high-temperature selective absorber coatings by physical vapor deposition (PVD) for solar thermal Application-A review. J. Alloys Compd. *839*, 155510.

Chang, J.-Y., Wang, H., and Wang, L. (2017). Tungsten Nanowire Metamaterials as Selective Solar Thermal Absorbers by Excitation of Magnetic Polaritons. J. Heat Trans. *139*, 052401-052408.

Chen, Y.-B., Zhang, Z.M., and Timans, P.J. (2006). Radiative Properties of Patterned Wafers With Nanoscale Linewidth. J. Heat Trans. *129*, 79-90.

Chen, Y.B., and Tan, K.H. (2010). The profile optimization of periodic nano-structures for wavelength-selective thermophotovoltaic emitters. Int. J. Heat Mass Trans. *53*, 5542-5551.

Enrico, A., Dubois, V., Niklaus, F., and Stemme, G. (2019). Scalable Manufacturing of Single Nanowire Devices Using Crack-Defined Shadow Mask Lithography. ACS Appl. Mater. Inter. *11*, 8217-8226.

Greiner, C.M., Iazikov, D., and Mossberg, T.W. (2006). Diffraction-limited performance of flat-substrate reflective imaging gratings patterned by DUV photolithography. Opt. Express *14*, 11952-11957.

Han, S., Shin, J.-H., Jung, P.-H., Lee, H., and Lee, B.J. (2016). Broadband Solar Thermal Absorber Based on Optical Metamaterials for High-Temperature Applications. Adv. Opt. Mater. *4*, 1265-1273.

Ho, C.-C., Chen, Y.-B., and Shih, F.-Y. (2016). Tailoring broadband radiative properties of glass with silver nano-pillars for saving energy. International Journal of Thermal Sciences *102*, 17-25.
Jae Lee, B., Chen, Y.-B., Han, S., Chiu, F.-C., and Jin Lee, H. (2014). Wavelength-Selective Solar Thermal Absorber With Two-Dimensional Nickel Gratings. J. Heat Trans. *136*, 072702-072707.

Jin, S., Lim, M., Lee, S.S., and Lee, B.J. (2016). Hyperbolic metamaterial-based near-field thermophotovoltaic system for hundreds of nanometer vacuum gap. Opt. Express *24*, A635-A649.

Khodasevych, I.E., Wang, L., Mitchell, A., and Rosengarten, G. (2015). Micro- and Nanostructured Surfaces for Selective Solar Absorption. Adv. Opt. Mater. *3*, 852-881.

Khoza, N., Nuru, Z.Y., Sackey, J., Kotsedi, L., Matinise, N., Ndlangamandla, C., and Maaza, M. (2019). Structural and optical properties of ZrOx/Zr/ZrOx/AlxOy multilayered coatings as selective solar absorbers. J. Alloys Compd. *773*, 975-979.

Li, P., Liu, B., Ni, Y., Liew, K.K., Sze, J., Chen, S., and Shen, S. (2015). Large-Scale Nanophotonic Solar Selective Absorbers for High-Efficiency Solar Thermal Energy Conversion. Adv. Mater. *27*, 4585-4591.

Lightsmyth http://www.lightsmyth.com/technology/ (accessed 21.8.20).

Lu, J.Y., Nam, S.H., Wilke, K., Raza, A., Lee, Y.E., AlGhaferi, A., Fang, N.X., and Zhang, T. (2016). Localized Surface Plasmon-Enhanced Ultrathin Film Broadband Nanoporous Absorbers. Adv. Opt. Mater. *4*, 1255-1264.

Lu, J.Y., Raza, A., Noorulla, S., Alketbi, A.S., Fang, N.X., Chen, G., and Zhang, T. (2017). Near-Perfect Ultrathin Nanocomposite Absorber with Self-Formed Topping Plasmonic Nanoparticles. Adv. Opt. Mater. *5*, 1700222.

Modest, M.F. (2013). Radiative heat transfer (Academic press).

Opachich, Y.P., Chen, N., Bell, P.M., Bradley, D.K., Feng, J., Gopal, A., Hilsabeck, T.J., Huffman, E., Koch, J.A., Landen, O.L.*, et al.* (2015). Precision fabrication of large area silicon-based geometrically enhanced x-ray photocathodes using plasma etching. Paper presented at: SPIE Optical Engineering + Applications (SPIE).

Park, Y.-S., and Lee, J.S. (2016). Location-Controlled Growth of Vertically Aligned Si Nanowires using Au Nanodisks Patterned by KrF Stepper Lithography. Chem. Asian J. *11*, 1878-1882.

Prasad, M.S., Mallikarjun, B., Ramakrishna, M., Joarder, J., Sobha, B., and Sakthivel, S. (2018). Zirconia nanoparticles embedded spinel selective absorber coating for high performance in open atmospheric condition. Sol. Energy Mater. Sol. *174*, 423-432.

Raman, A.P., Anoma, M.A., Zhu, L., Rephaeli, E., and Fan, S. (2014). Passive radiative cooling below ambient air temperature under direct sunlight. Nature *515*, 540.

Sanjay, K.S., Dinesh, K., Schmitt, S.W., Sood, K.N., Christiansen, S.H., and Singh, P.K. (2014). Large area fabrication of vertical silicon nanowire arrays by silver-assisted single-step chemical etching and their formation kinetics. Nanotechnology *25*, 175601.

Seo, J., Jung, P.-H., Kim, M., Yang, S., Lee, I., Lee, J., Lee, H., and Lee, B.J. (2019). Design of a Broadband Solar Thermal Absorber Using a Deep Neural Network and Experimental Demonstration of Its Performance. Sci. Rep. *9*, 15028.

Smith, D., Shiles, E., Inokuti, M., and Palik, E. (1985). Handbook of Optical constants of Solids (Orlando: Academic press).

Thomas, N.H., Chen, Z., Fan, S., and Minnich, A.J. (2017). Semiconductor-based Multilayer Selective Solar Absorber for Unconcentrated Solar Thermal Energy Conversion. Sci. Rep. *7*, 5362.

Tian, Y., Ghanekar, A., Ricci, M., Hyde, M., Gregory, O., and Zheng, Y. (2018). A Review of Tunable Wavelength Selectivity of Metamaterials in Near-Field and Far-Field Radiative Thermal Transport. Materials *11*, 862.

Wang, H., Alshehri, H., Su, H., and Wang, L. (2018). Design, fabrication and optical characterizations of large-area lithography-free ultrathin multilayer selective solar coatings with excellent thermal stability in air. Sol. Energy Mater. Sol. *174*, 445-452.

Wang, H., Prasad Sivan, V., Mitchell, A., Rosengarten, G., Phelan, P., and Wang, L. (2015). Highly efficient selective metamaterial absorber for high-temperature solar thermal energy harvesting. Sol. Energy Mater. Sol. *137*, 235-242.

Wang, H., and Wang, L. (2013). Perfect selective metamaterial solar absorbers. Opt. Express *21*, A1078-A1093.

Woolf, D.N., Kadlec, E.A., Bethke, D., Grine, A.D., Nogan, J.J., Cederberg, J.G., Bruce Burckel, D., Luk, T.S., Shaner, E.A., and Hensley, J.M. (2018). High-efficiency thermophotovoltaic energy conversion enabled by a metamaterial selective emitter. Optica *5*, 213-218.

Xu, K., Du, M., Hao, L., Mi, J., Yu, Q., and Li, S. (2020). A review of high-temperature selective absorbing coatings for solar thermal applications. J. Materiomics *6*, 167-182.

Zhai, Y., Ma, Y., David, S.N., Zhao, D., Lou, R., Tan, G., Yang, R., and Yin, X. (2017). Scalable-manufactured randomized glass-polymer hybrid metamaterial for daytime radiative cooling. Science *355*, 1062-1066.

Zhang, Z.M. (2007). Nano/microscale heat transfer (McGraw-Hill New York).

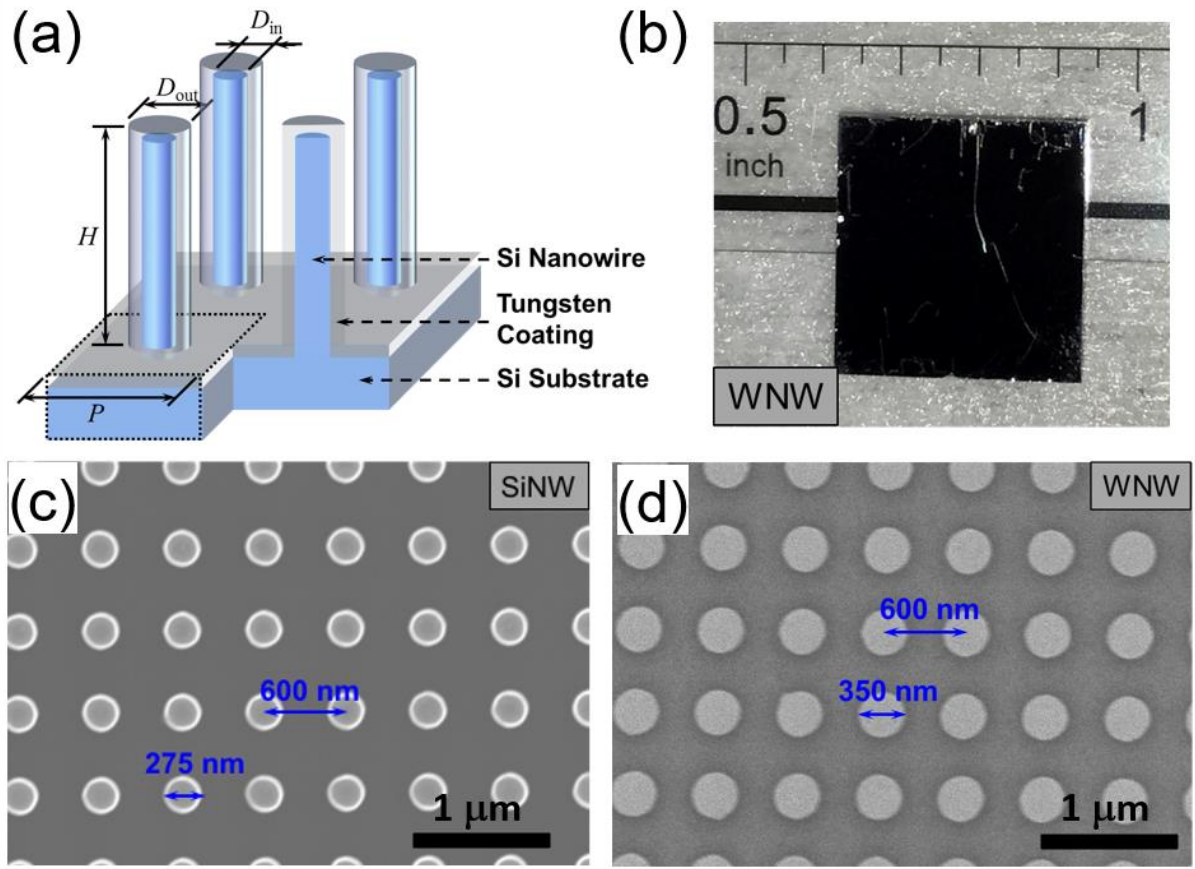

**Figure 1.** (a) Structural schematic of Si-cored tungsten nanowire (WNW) metamaterial as selective solar absorber where a layer of tungsten is conformably coated on the silicon nanowire (SiNW) surfaces. (b) Photo of fabricated Si-cored WNW metamaterial sample in a size of 8×8 mm$^2$ with visibly black appearance. (c) Top-view SEM image of bare SiNW array with a period of 600 nm and nanowire diameter of 275 nm. (d) Top-view SEM image of Si-cored WNW array with the same period of 600 nm and increased nanowire diameter of 350 nm after 37.5-nm tungsten is conformally sputtered on the silicon surfaces.

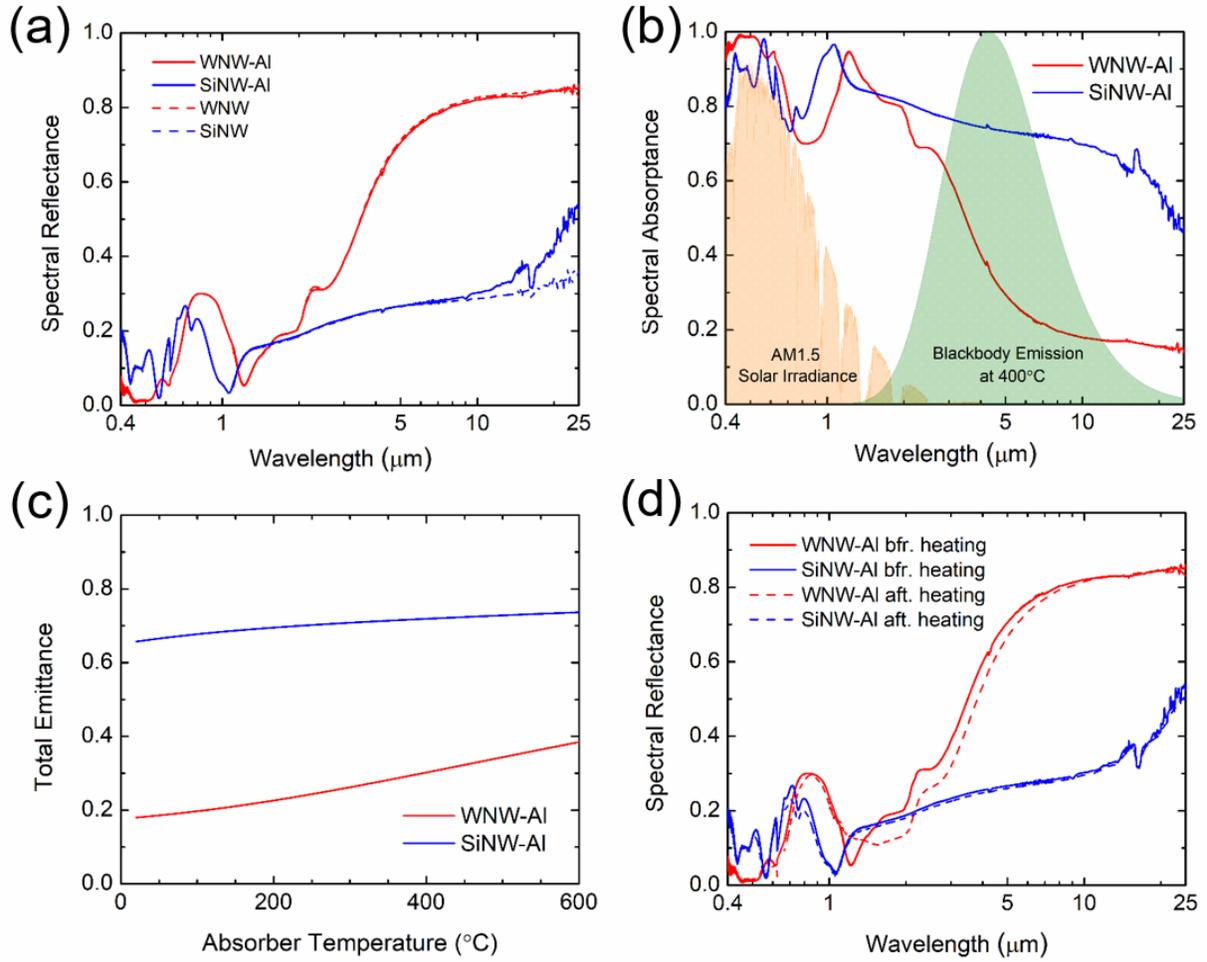

**Figure 2.** (a) Measured spectral (near normal) reflectance ($R_\lambda$) in the broad spectral range from 0.4 μm to 25 μm for bare silicon nanowire (SiNW) and Si-cored tungsten nanowire (WNW) before and after the deposition of 200-nm-thick Al on the backside of samples. (b) Spectral absorptance (or spectral emittance $\varepsilon_\lambda = \alpha_\lambda$ according to Kirchhoff's law) obtained from measured spectral reflectance (i.e., $\alpha_\lambda = 1 - R_\lambda$) in the broad spectral range from visible to mid-infrared for opaque SiNW and Si-cored WNW with Al backside coating (denoted as SiNW-Al and WNW-Al). The shaded areas represent the spectral solar irradiance from air mass 1.5 Global tilt (orange color) and spectral blackbody emissive power at 400°C (green color) both normalized to the peak spectral emissive power of 0.178 W/cm$^2$/μm. (c) Calculated total emittance of WNW-Al and SiNW-Al as a function of absorber temperature. (d) Measured spectral reflectance of WNW-Al and SiNW-Al samples in the broad spectral range at room temperature before and after the solar thermal heating tests in vacuum with a maximum temperature of 375°C for 10 hrs.

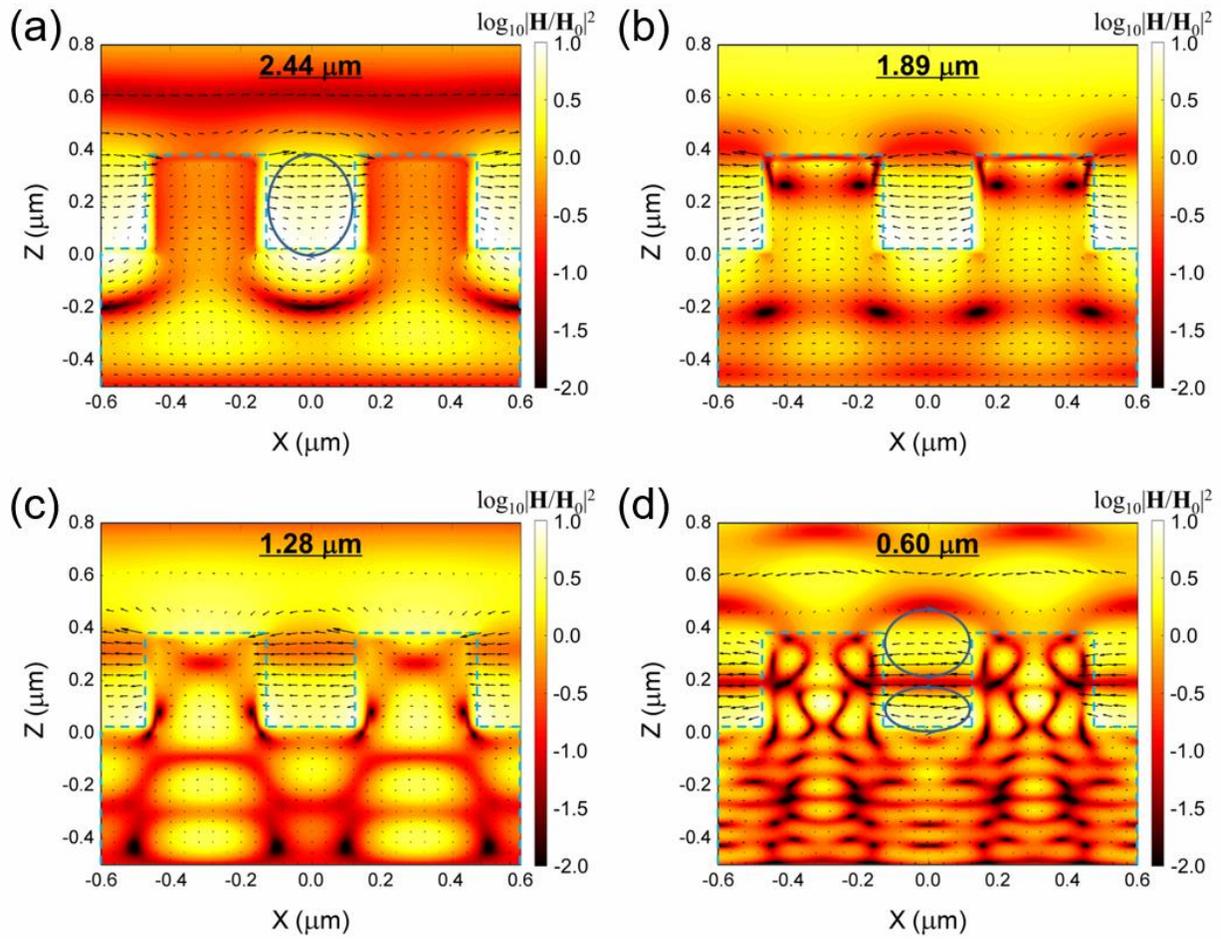

**Figure 3.** Numerical simulations of electromagnetic fields distribution for tungsten nanowire arrays with the same geometry (i.e., 600 nm period, 350 nm diameter, 350 nm height) at selected wavelengths of (a) 2.44 μm, (b) 1.89 μm, (c) 1.28 μm, and (d) 0.60 μm where absorptance peaks are observed from the measured spectra. Color contour presents the local magnetic field strength, while the arrows indicates the electric field vectors.

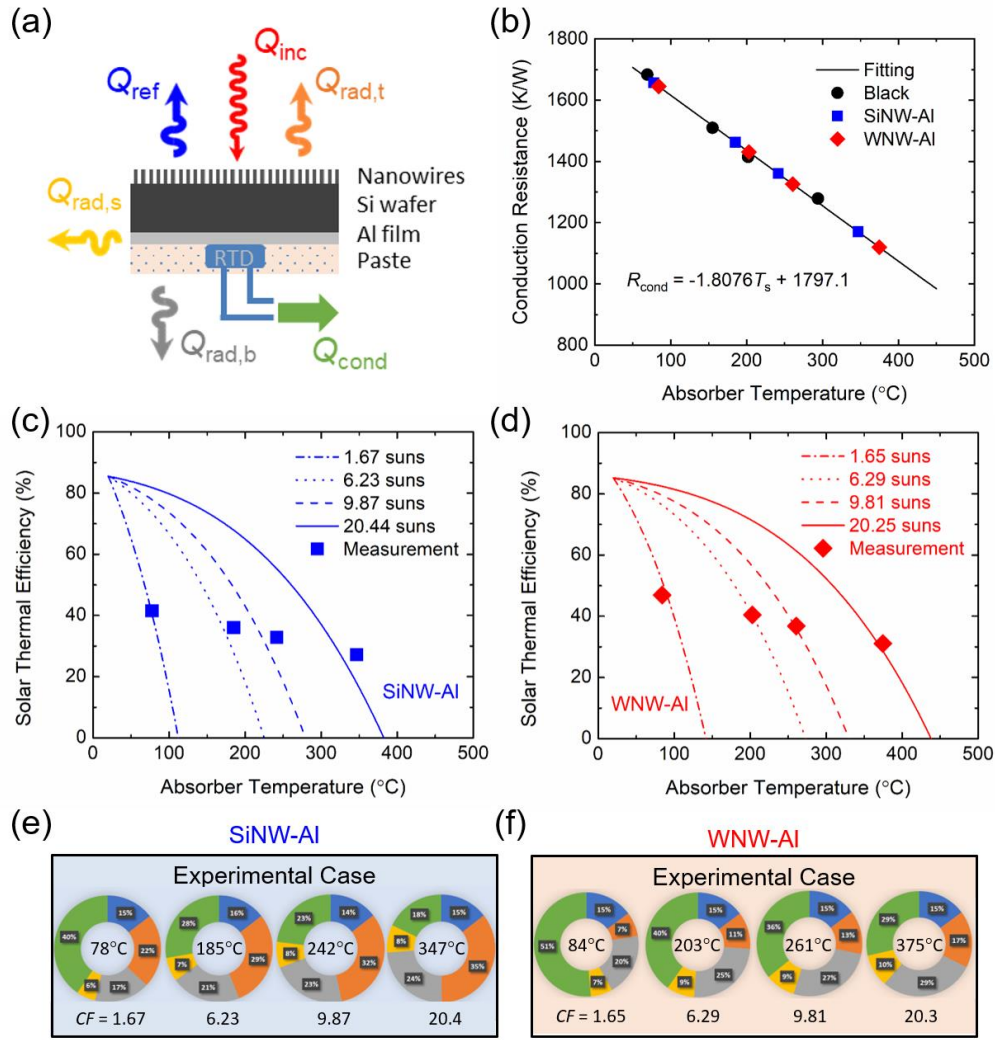

**Figure 4.** (a) Schematic of heat transfer modes for nanowire absorbers during the lab-scale solar thermal testing at different solar concentrations. Note that the convective heat loss is neglected as the test was conducted under high vacuum environment, and the absorber temperature was measured by a resistance temperature detector (RTD) glued at the backside with thermal paste. (b) Calibration of conduction resistance for the conduction heat loss via the RTD wires (considered as a useful heat gain) at multiple absorber temperatures with a black absorber, based on which the conduction heat gain is calculated for nanowire samples from the measured temperatures. (c-d) Measured solar thermal efficiency (markers) at multiple solar concentrations (from ~1.6 to ~20 suns) for fabricated SiNW-Al and WNW-Al absorber samples in good agreement with the theoretical prediction (lines). (e-f) Heat transfer analysis pie charts for illustrating the energy loss ratios for SiNW-Al and WNW-Al samples at measured absorber temperatures under corresponding solar concentration factors (CF).

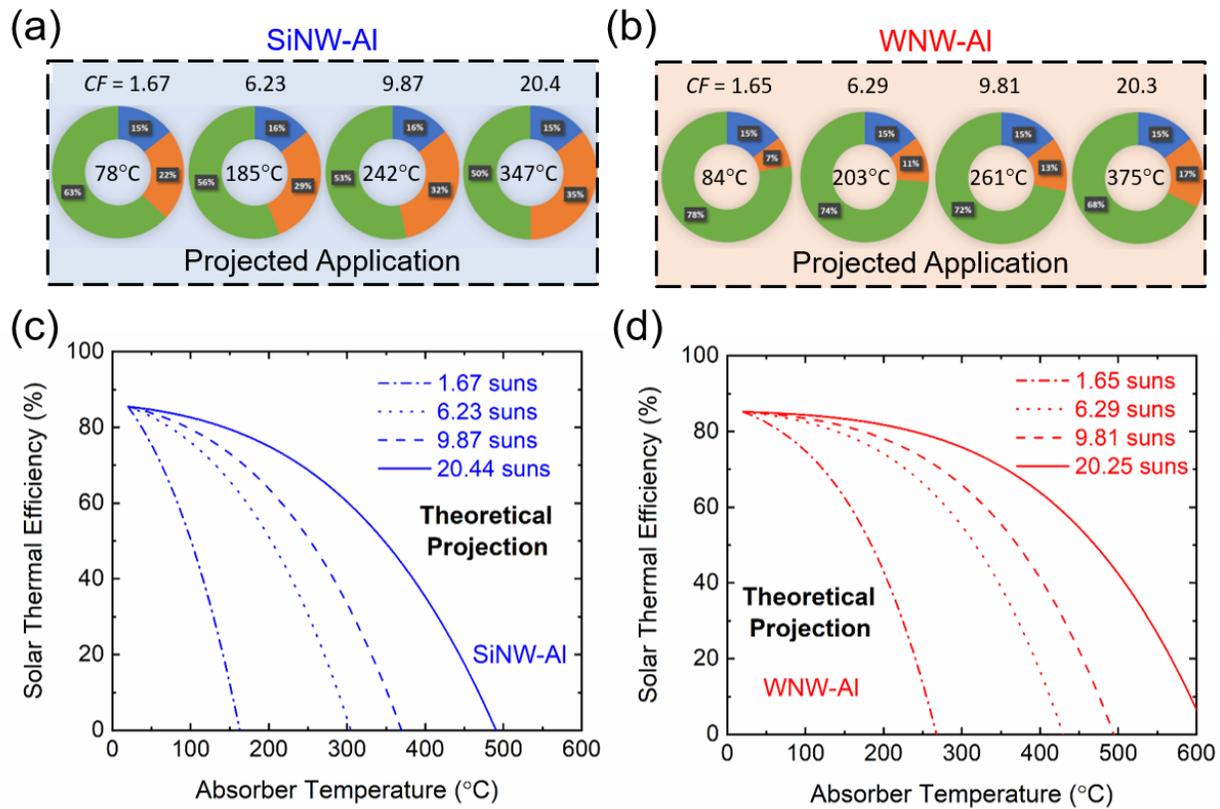

**Figure 5.** (a-b) Heat transfer analysis pie charts for illustrating the energy loss ratios for SiNW-Al and WNW-Al samples at measured absorber temperatures under corresponding solar concentration factors (CF), and (c-d) theoretically predicted solar thermal efficiency at multiple solar concentrations for fabricated SiNW-Al and WNW-Al samples as a function of absorber temperature for projected solar thermal applications where the radiation losses from the side and bottom surfaces can be eliminated.